\documentclass{amsproc}
\newtheorem{thm}{Theorem}[section]

\newtheorem{lem}[thm]{Lemma}

\theoremstyle{definition}
\newtheorem{defn}[thm]{Definition}
\theoremstyle{remark}
\newtheorem{rem}[thm]{Remark}
\numberwithin{equation}{section}
\usepackage{tikz}

\begin{document}
\title{Tropical geometric interpretation of ultradiscrete singularity confinement}%
\author{Christopher M. Ormerod}%
\address{Department of Mathematics and Statistics, La Trobe University, Bundoora, Vic 3086, Australia.}%
\email{christopher.ormerod@gmail.com}%

\begin{abstract}
Using the interpretation of the ultradiscretization procedure as a non-Archimedean valuation, we use results of tropical geometry to show how roots and poles manifest themselves in piece-wise linear systems as points of non-differentiability. This will allow us to demonstrate a correspondence between singularity confinement for discrete integrable systems and ultradiscrete singularity confinement for ultradiscrete integrable systems. 
\end{abstract}


\maketitle

\section{Introduction}

The ultradiscretization procedure takes a subtraction-free rational function to a function defined over the max-plus semifield via a non-analytic limit \cite{ultradiscretization}. This procedure famously linked integrable soliton equations to integrable cellular automata, hence, the ultradiscretization procedure was thought to preserve properties associated with integrability \cite{udKP:BBS, udmKdV:BBS, ultradiscretization}. The ultradiscretization procedure has been used to create various new and novel ultradiscrete systems such as the ultradiscrete Painlev\'e equations \cite{ultimate, Ormerod:qP6}, ultradiscrete Quispel-Roberts-Thompson maps \cite{NobeQRT} and ultradiscrete lattice equations \cite{ultradiscretization}.

Many of the key properties associated of the original systems seemed to survive the ultradiscretization procedure; the Lax integrability \cite{Quispel:UDLax, Ormerod:uP3}, symmetries \cite{YamadaEreps}, conserved quantities \cite{UDconserved} and special solutions \cite{Ormerod:uhypergeometric}. Recently, the interesting observation was made by Hone and Fordy that the ultradiscrete versions of mappings govern the growth of degrees of (subtraction-free) mappings in the initial conditions \cite{fordy2012discrete}.

The crucial point this note makes is that one may interpret the ultradiscretization procedure as a map that send points singular points, such as poles and roots, to points of non-differentiability. This involves considering the ultradiscretization of arbitrary rational functions, not just subtraction-free rational functions. With this in mind, the ultradiscrete singularity confinement proposed by La Fortune and Joshi \cite{Joshi:TropSC} may be understood as an image of singularity confinement \cite{Gramani:DPainleveProperty, Gramani:DiscretePs}. We have attempted to capture the essence of the argument with a number of simple instructive examples. 

For this to make sense, we first interpret the ultradiscretization procedure as a non-Archimedean valuation, which allows the ultradiscretization procedure to be understood in terms of tropical geometry \cite{TropicalGeometry}. By realizing the ultradiscretization procedure as a non-Archimedean valuation, the image of a singularity makes sense. This will also allow us utilize the results of tropical geometry \cite{FieldforTG}, or more precisely, the original results of Bieri and Groves \cite{BieriGroves} and Richter-Gebert et al. \cite{TropicalGeometry}, to show how roots and poles map to points of non-differentiability. One of the implications of this interpretation is a correspondence between singularity confinement for discrete integrable systems and the proposed ultradiscrete version of singularity confinement \cite{Joshi:TropSC} over the max-plus semifield \cite{pin1998tropical}.

There have been many attempts at reconciling the role of subtraction in the ultradiscretization procedure, such as the so-called inversible max-plus algebra \cite{Ochiai}, the $s$-ultradiscretization \cite{s-ultradiscretization} and more analytic approaches \cite{Lafortune:negativity}. The approach that is closest to our framework is that of Kasman and Lafortune \cite{Lafortune:negativity}, yet they do not make any tropical geometric connections in their work. Algebraically, this approach is actually closer to the arithmetic integrability of Kanki et al. \cite{DPIIFiniteField}. 

This brings us to our second underlying point; we wish to strengthen the argument that a fundamental understanding of the geometry of these piece-wise linear systems should be phrased in terms of the points of non-differentiability. There are some who believe that the concept of a singularity is lost when passing to the ultradiscrete, however, we argue through the realisation of the ultradiscretization procedure as a non-Archimedean valuation that the singularities have just manifested themselves in a different way.

In section \S \ref{framework}, we will introduce the ultradiscretization procedure and show how to realise the ultradiscretization procedure as a lift to an extended version of the algebraic functions followed by the application of a non-Archimedean valuation. In \S \ref{ODEs} we will study some particular cases of both ultradiscrete ordinary difference and partial difference equations to see how using the valuations around various singularities of the lifted equations map to points of discontinuity of the ultradiscrete systems. 

\section{Ultradiscretization}\label{framework}

The most basic algebraic domain of the ultradiscrete equations are the tropical semirings, which are the $\min$-plus and $\max$-plus semirings (which are isomorphic to each other) \cite{pin1998tropical}. From an integrable perspective, we will use the max-plus semifield, which we set to be $\mathbb{S} = \mathbb{R} \cup \{-\infty\}$, with operations of $\otimes$ and $\oplus$, defined as
\begin{align*}
a \otimes b &= a+ b,\\
a \oplus b &= \max(a,b),
\end{align*}
referred to as tropical multiplication and tropical addition respectively (with a natural tropical division) \cite{TropicalGeometry}. The element $-\infty$ is a tropical additive identity and is adjoined for convenience. These operations may be extended to matrices over the max-plus semiring in a natural way  \cite{Quispel:UDLax, Ormerod:uP3} and also to the field of tropically rational functions, $\mathbb{S}[X_1,\ldots, X_n]$, in a natural manner. The reason this is a semifield is that there is a tropical division, $\oslash$, but there is no tropical analogue of the subtraction operation \cite{TropicalGeometry}. 

\subsection{Ultradiscretization as a non-analytic limit}

Let us first recount the definition of the ultradiscretization procedure as it appears in the literature \cite{ultradiscretization}. Let us start with a rational function of a number of positive variables, $f(x_1,\ldots, x_n)$, which does not, require the operation of subtraction to be expressed (i.e., the function is subtraction-free). We introduce a number of ultradiscrete variables, $X_i$, via the relation $x_i = e^{X_i/\epsilon}$. The ultradiscretization of $f$, denoted $F(X_1, \ldots, X_n)$, is defined as the non-analytic limit
\begin{equation}\label{eq1:ultradiscretization}
F(X_1, \ldots, X_n) := \lim_{\epsilon \to 0} \epsilon \log f\left( e^{X_1/\epsilon}, \ldots, e^{X_n/\epsilon} \right).
\end{equation}
The positivity of the variables and subtraction-free nature of the function plays a crucial role in making the ultradiscretization procedure a homomorphism of the semifield of subtraction-free functions to the semifield, $\mathbb{S}[X_1, \ldots, X_n]$.  The way in which this homomorphism may be computed is simply via the following replacement of binary operations:
\begin{subequations}\label{eq2:equiv}
\begin{align}
\label{eq2:equiva}x_1 + x_2 &\to \max(X_1,X_2) = X _1 \oplus X_2,\\
\label{eq2:equivb}x_1 x_2 &\to X_1 + X_2 = X_1 \otimes X_2.
\end{align}
\end{subequations}

\begin{lem}\label{lem2:equiv}
The ultradiscretization defined by \eqref{eq1:ultradiscretization} of any (subtraction-free) rational function is equivalent to the replacement of the $x_i$ with $X_i$ and the replacement of binary operations defined by \eqref{eq2:equiv}.
\end{lem}

To obtain an ultradiscrete system, we take a known equation that defines a discrete integrable system and perform the ultradiscretization procedure to both sides. For example, let us take the QRT mapping defined by the recurrence relation
\begin{equation}\label{QRT}
x_{n-1}x_{n+1} = \dfrac{a_3a_4 (x_n+a_1)(x_n+a_2)}{(x_{n}+a_3)(x_{n}+a_4)},
\end{equation}
which is related to the map
\[
\phi \begin{pmatrix} x \\ y \end{pmatrix} = \begin{pmatrix} y \\ \dfrac{a_3a_4 (y+a_1)(y+a_2)}{x(y+a_3)(y+a_4)}\end{pmatrix},
\]
which preserves the biquadratic
\[
I_n = \frac{\left(a_1+y\right) \left(a _2+y\right)}{x y}+\frac{x \left(a _3+y\right) \left(a_4+y\right)}{a_3 a_4 y}+\frac{a_1+a_2+\left(\frac{1}{a_4}+\frac{1}{a_3}\right) y^2}{y}.
\]
To obtain the systems ultradiscretization, we apply the procedure to the left-hand-side and right-hand-side. This gives the ultradiscrete analogue of \eqref{QRT} as
\begin{align}\label{uP3}
X_{n-1} + X_{n+1} = A_3 + A_4 &+ \max(X_n, A_1) + \max(X_n,A_2)\\
\nonumber &- \max(X_n,A_3) - \max(X_n,A_4) ,
\end{align}
which preserves the tropical biquadratic
\begin{align*}
I_n = \max(& \max(Y,A_1) + \max(Y,A_2)- X-Y, \max(A_1,A_2)-Y,\\
& Y-\min(A_3,A_4), X-Y+\max(Y-A_3,0)+\max(Y-A_4,0) ).
\end{align*}
The resulting system may be defined over the integers, hence, these systems are often referred to as cellular automata \cite{Ormerod:uP3}. 

\subsection{Ultradiscretization as a non-Archimedean valuation}

Tropical geometry is a skeletonization of algebraic geometry where the main objects of interest are convex polyhedra \cite{TropicalGeometry}. A tool that is ubiquitous in this theory is the realisation of the image of a variety under a non-Archimedean valuation as a convex polyhedra, which was initially established by Bieri and Groves \cite{BieriGroves} and Richter-Gebert et al. \cite{TropicalGeometry}. To elucidate the link between tropical geometry and the ultradiscretization process, we must first realise the ultradiscretization procedure as a non-Archimedean valuation. 

The base field that is often used in tropical geometry is the field of Puiseux series \cite{TropicalGeometry}. Let us first consider the field of Puiseux series in an indeterminant, $t$, given by
\[
\mathbb{K} = \bigcup_{n = 1}^{\infty} \mathbb{C}\left(\left( t^{\frac{1}{n}} \right)\right).
\]
Every element, by definition, may be expressed in terms of 
\[
f(t) = a_0 t^{q_0} + a_1 t^{q_1} + \ldots
\]
where $a_0 \neq 0$ and the $q_i \in \mathbb{Q}$ are ordered such that $q_i < q_{i+1}$ for all $i$, and that the set $q_i$ contains no accumulation points (well ordered) \cite{TropicalGeometry}. The Puiseux series over $\mathbb{C}$ is an algebraically closed field of characteristic $0$. It may also be thought of as the algebraic closure of the field of Laurent series. There is a natural valuation on this field,
\begin{equation}\label{valuation}
\nu\left(a_0 t^{q_0} + \ldots \right) = -q_0.
\end{equation}
In general, $\nu$ satisfies the properties
\begin{enumerate}
\item[1.]{$\nu(f) = -\infty$ if and only if $f = 0$.}
\item[2.]{$\nu(fg) = \nu(f)+\nu(g)$.}
\item[3.]{$\nu(f+g) \leq \nu(f) + \nu(g)$.}
\end{enumerate}
which implies $\nu$ is a valuation. However, $\nu$ also satisfies the following additional non-Archimedean triangle inequality:
\begin{enumerate}
\item[3a.]{$\nu(f+g) \leq \max(\nu(f),\nu(g))$.}
\end{enumerate}
Any valuation field is metrizable, where the metric is
\begin{equation}\label{metric}
d(f,g) = e^{\nu(f-g)}.
\end{equation}
The Puiseux series have two minor deficiencies; firstly that the valuation group (i.e., the image of $\nu$) is necessarily a rational number, hence, tropical geometers require an additional closure step in their framework, and secondly that the field is not complete with respect to the metric. One way to overcome this is to deal with a more general field, which comprises of the formal series expansions in an indeterminant $t$ whose powers are elements of $\mathbb{R}$. One may find a suitable field extension, from which one may lift the valuation, to a valuation field that is algebraically closed and complete;
\[
\overline{\mathbb{K}} = \left\{ f = \sum_{i} a_i t^{q_i} | q_i \textrm{ are well ordered.}\right\}
\]
This was introduced in \cite{FieldforTG}, however, an isomorphic algebraic construction appeared in \cite{Ormerod:theory}. The valuation is still defined by \eqref{valuation}, and the metric is still given by \eqref{metric}. This field also has the advantage that it's valuation group is $\mathbb{R}$, making it an excellent candidate as a base field for the use in tropical geometry \cite{FieldforTG}. To express the ultradiscretization as a non-Archimedean valuation, for every positive variable, $x_i$, we introduce ultradiscrete variables, $X_i$, via the relation $x_i = t^{-X_i}$. Now the ultradiscretization of a subtraction-free function $f(x_1,\ldots, x_n)$, denoted $F(X_1,\ldots, X_n)$, is now simply defined as
\begin{equation}\label{eq2:ultradiscretization}
F(X_1, \ldots, X_n) := \nu\left( f\left(t^{-X_1}, \ldots, t^{-X_n} \right) \right).
\end{equation}

\begin{lem}
The ultradiscretization procedures defined by \eqref{eq1:ultradiscretization} and \eqref{eq2:ultradiscretization} coincide.
\end{lem}

One of the advantages of viewing the ultradiscretization in this way is that the valuation is defined for all rational functions. It still presents a problem in that the ultradiscretization procedure is not a homomorphism of semirings, as for arbitrary rational functions, the equality in the ultradiscretization procedure is replaced by an inequality. We are not at a total loss, as we have at our disposal a well known simple lemma.

\begin{lem}\label{equal}
If $\nu(f) \neq \nu(g)$, then $\nu(f + g) = \max(\nu(f),\nu(g))$. 
\end{lem}

We used this lemma to show that a hypergeometric solution over $\mathbb{K}$ faithfully mapped to a hypergeometric solution of $u$-$\mathrm{P}_{III}$ over $\mathbb{S}$ \cite{Ormerod:uhypergeometric}. We now state the main theorem that we will use.

\begin{thm}
Given a principle ideal, $I = \langle f \rangle \subset \mathbb{K}(x_1, \ldots, x_n)$, then the closure of the image, $\mathrm{Trop}(I) = \overline{\nu(V(I))}$, coincides with the points where the ultradiscretization of $f$ is not differentiable. 
\end{thm}

Despite the integrable wording, this is a result that should be rightly attributed to those in tropical geometry \cite{BieriGroves, TropicalGeometry, FieldforTG}. We have assumed a bit of notation here, where $V(I) \subset \mathbb{K}^n$, from algebraic geometry, is the variety associated with an ideal:
\[
V(I) = \bigcap_{g \in I} \left\{ g(x) = 0, x \in \mathbb{K} \right\},
\]
and the $\overline{S}$, for any set $S$, is the topological closure with respect to the Euclidean metric on $\mathbb{R}^n$. Naturally, if we deal with a field whose valuation group is $\mathbb{R}$ itself (rather than $\mathbb{Q}$), then this topological closure is not necessary \cite{FieldforTG}, however, we include it for the correspondence with the more standard work on tropical geometry \cite{TropicalGeometry}. One final result required for this correspondence to be complete is that the tropical variety is a function only of the valuations of the various coefficients \cite{Amoebas}. This correspondence provides tropical geometers a straightforward way of computing the tropical variety \cite{TropicalGeometry}.

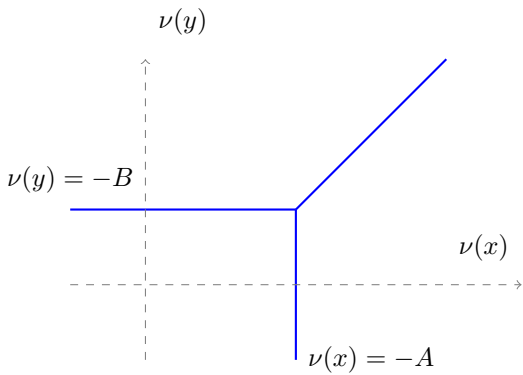
\begin{figure}[!ht]
\begin{tikzpicture}
\draw[thick,blue] (0,2) -- (3,2);
\draw[thick,blue] (3,2) -- (5,4);
\draw[thick,blue] (3,0) -- (3,2);
\draw[->,dashed,color=black!50] (0,1) -- (6,1);
\node at (5.5,1.5) {$\nu(x)$};
\draw[->,dashed,color=black!50] (1,0) -- (1,4);
\node at (1.5,4.5) {$\nu(y)$};
\node at (4,0) {$\nu(x) = -A$};
\node at (0,2.4) {$\nu(y) = -B$};
\end{tikzpicture}
\caption{A tropical degree 1 curve in $\mathbb{R}^2$.\label{line}}
\end{figure}

Let us demonstrate this by considering a linear system
\[
f(x,y) = ax+by+1.
\]
Let us lift this to the field, $\mathbb{K}$, in a variable, $t$, in which the valuations of $a$ and $b$ are $\nu(a) = A$ and $\nu(b) = B$. By setting $f(x,y) = 0$, we may write $y$ as a function of $x$, or vise versa, hence, 
\[
y =  -\dfrac{c+ax}{b},
\]
in which, by applying the valuation, and using lemma \ref{equal}, if $\nu(x)+A \neq 0$, then $\nu(x) = \max(0, A+ \nu(y)) - B$. Conversely, expressing $x$ as a function of $y$, we obtain $\nu(x) = \max(0, B+\nu(y)) - A$. This allows us to determine $\mathrm{Trop}(\langle f \rangle)$, which we characterize as a node at $(-A,-B)$ with three semi-infinite rays, as depicted in figure \ref{line}.
As for what the correspondence is with the ultradiscretization procedure, since $f$ is subtraction free, we apply our valuation, where $\nu(x) = X$ and $\nu(y) = Y$, which are assumed to be subtraction free, to give an ultradiscretized function
\[
F(X,Y) := \max(A+X,Y+B,0).
\]
This defines three regions of $\mathbb{R}^2$ defined by which of the three arguments of the $\max$ is dominant, and the points in which there is a switch between one of the arguments being dominant to another argument being dominant is clearly a point in which $F$ is not differentiable. The remainder is easy to see. 

In addition to the this correspondence, one may also compute the stable intersection\footnote{For a definition of stable intersection, we refer to \cite{TropicalGeometry}.} of two lines fairly easily from this framework. Let us consider another line over $\mathbb{K}$, given by $cx+dy+1 = 0$. It may be associated with a tropical line defined by the points of non-differentiability of 
\[
G(X,Y) := \max(C+X,D+Y,0),
\]
where $\nu(c) = C$ and $\nu(d) = D$. If we wish to find the stable intersection of the tropical lines, we find that the intersection over $\mathbb{K}$ is given by
\[
p = \left( \dfrac{b-d}{a d-b c},\dfrac{a-c}{b c-a d}\right),
\]
whose valuation, under the assumption that 
\[
(\nu(b)-\nu(d))(\nu(a)-\nu(c))(\nu(ad) - \nu(bc)) \neq 0,
\]
gives us the (stable) intersection point of the tropical lines defined by $F$ and $G$ as
\[
P = \left( \max(B,D) - \max(A+D,B+C) , \max(A,C) - \max(A+D,B+C) \right).
\]
One can verify this statement simply using elementary arguments, however, the point is that the point is easily calculable from using $\mathbb{K}$ \cite{TropicalGeometry}. 

\section{Singularity Confinement}\label{ODEs}

The singularity confinement property for maps has been used in many contexts; to de-autonomize known integrable autonomous maps \cite{Gramani:DiscretePs}, to define new integrable interesting mappings and to distinguish integrable maps from non-integrable maps \cite{Gramani:DPainleveProperty}. Despite some failings (see \cite{Viallet}) it remains very useful in defining connections between the theory of integrable mappings and rational surfaces \cite{Sakai:Rational, Duistermaat}. For lattice equations, the singularity confinement property has been used to provide evidence for integrability \cite{Gramani:DPainleveProperty} and to deautonomize lattice equations \cite{Gramani:ABSSC}.

\subsection{Tropical Singularity confinement for O$\Delta$E's}

In terms of the definition of singularity confinement, the original presentation was via a series of examples. This definition can be made more concrete if we are able to phrase it in terms of injectivity. Let us consider a map, $\phi : \mathbb{C}^n \to \mathbb{C}^n$, which defines a discrete dynamical system by letting $x_{k+1} = \phi(x_k)$. A point is called a singularity if it is not injective, more to the point, the determinant of the Jacobian is $0$.

\begin{defn}
We say that a map has the singularity confinement property if for all $x$ such that $\phi$ is singular at $x$, there exists an $n$ such that $x$ is removable singularity of $\phi^n$.
\end{defn}

Let us consider a class of QRT maps that are specified by the recurrence relation
\begin{equation}\label{QRT1}
x_{n+1} x_n^\sigma x_{n-1} = a + b x_n,
\end{equation}
which is associated with the mapping
\[
\phi \begin{pmatrix} x \\ y \end{pmatrix} = \begin{pmatrix} y \\ \dfrac{a+by}{x y^\sigma} \end{pmatrix}, 
\]
where $\sigma = 0,1,2$ and $3$. This is a equation that has been considered numerous times to single out integrable equations from non-integrable ones. When $\sigma = 0,1$ or $2$, this is an integrable difference equation, in fact, they preserve the integrals
\begin{align*}
I_0 =& \frac{y \left(a+b^2+x^2\right)+(b+x) (a+b x)+y^2 (b+x)}{b x y},\\
I_1 =& \frac{a+b x+b y+x^2 y+x y^2}{x y},\\
I_2 =& \frac{a+b x+b y+x^2 y^2}{x y},
\end{align*}
respectively, whereas the case in which $\sigma = 3$ is not integrable. In each case, if we let $x_1 = -a/b + \epsilon$, then 
\[
\lim_{\epsilon \to 0} x_2 = 0.
\]
As a map in $(x_{k-1},x_k)$-space, this sends a line, parameterized by $x_0$, to a single point, $(-a/b,0)$. In the limit as $\epsilon \to 0$, the iterates are
\[
\begin{array}{c||c|c|c|c|c|c|c|c|c|}
\sigma  & x_1 & x_2 & x_3 & x_4 & x_5 & x_6 & x_7 & x_8 & x_9\\ \hline \hline
0 &  -\frac{a}{b} & 0 & -b & \infty & \infty & -b & 0 & -\frac{a}{b} & x_0 \\
1 &  -\frac{a}{b} & 0 & \infty & \infty & 0 & -\frac{a}{b} & x_0 & * & * \\
2 &  -\frac{a}{b} & 0 & \infty & 0 & -\frac{a}{b} & x_0 & * & * & *\\
3 &  -\frac{a}{b} & 0 & \infty & 0 & \infty & 0 & \infty & 0 & \infty
\end{array}.
\]
We readily find that one of the iterates for the integrable cases contains the data from the initial condition, $x_0$ (we have omitted iterates after that point). The non-integrable case alternates between $(x_{k-1},x_{k}) = (0,\infty)$ and $(\infty,0)$ ad infinitum. Another way to calculate this singularity pattern, which is relevant to the correspondence between singularity confinement and tropical singularity confinement, is to start with $x_1 = -b/a e^{\delta}$ where $\delta \to 0$. The calculations follow analogously.

The above constitutes the way in which this procedure is usually viewed \cite{Gramani:DPainleveProperty}. Alternatively, we incorporates the definition more directly; one may calculate the determinant of Jacobian of $\phi^n$ in the limit as $y \to -a/b$ (x is free) for each of the maps for successive values of $n$, finding readily that the determinants are non-zero (in fact, they are $1$) when $n = 8$ for $\sigma =0$ \footnote{When $a = b = 1$, the third power of the mapping is regular.  i.e., the singularity is confined earlier than $n=8$ in the special case when $a=b=1$.}, $n=6$ for $\sigma = 1$ and $n=5$ for $\sigma = 2$.

We may now turn to the tropical analogue of singularity confinement, as discussed by Joshi and Lafortune \cite{Joshi:TropSC}.

\begin{defn}
We say that a piece-wise linear map has the tropical singularity confinement property if for all $x$ such that $\Phi$ is not differentiable at $x$, there exists an $n$ such that $\Phi^n$ that is differentiable at $x$.
\end{defn}

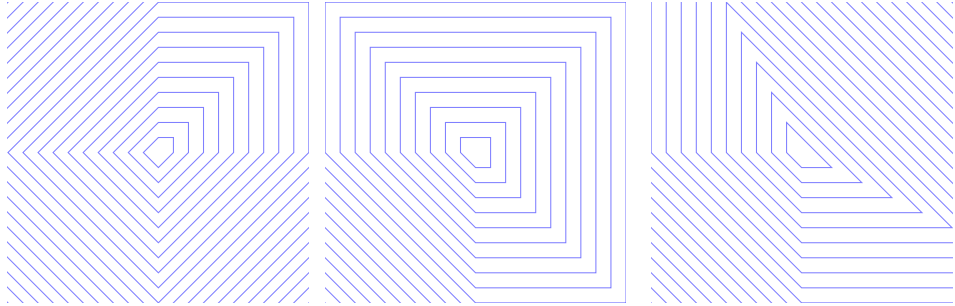
\begin{figure}[!ht]
\begin{tikzpicture}[scale=1]
\clip(-2,-2) -- (2,-2) -- (2,2)--(-2,2);
\clip(-2,-2) -- (2,-2) -- (2,2)--(-2,2);
\foreach \x in {0,.2,...,5}
{
	\draw[blue!50](-\x,0) -- (0,-\x)--(\x,0) -- (\x,\x)--(0,\x)--cycle;
}
\end{tikzpicture}\hspace{.1cm}
\begin{tikzpicture}[scale=1]
\clip(-2,-2) -- (2,-2) -- (2,2)--(-2,2);
\foreach \x in {0,.2,...,5}
{
	\draw[blue!50](-\x,0) -- (0,-\x)--(\x,-\x) -- (\x,\x)--(-\x,\x)--cycle;
}
\end{tikzpicture} \hspace{.1cm}
\begin{tikzpicture}
\clip(-2,-2) -- (2,-2) -- (2,2)--(-2,2);
\foreach \x in {0,.2,...,5}
{
	\draw[blue!50](-\x,0) -- (0,-\x)--(2*\x,-\x) -- (-\x,2*\x)--cycle;
}
\end{tikzpicture}
\caption{The level sets for the invariants, $I_0$, $I_1$ and $I_2$, of \eqref{QRT1} where, from left to right, $\sigma= 0$, $1$ and $2$ respectively.\label{levelsets}}
\end{figure}

Let us take the tropical analogue of \eqref{QRT1}, which is
\begin{equation}\label{uQRT}
X_{n-1} + \sigma X_n + X_{n-1} = \max(A, B+ X_n),
\end{equation}
which is in the class of ultradiscrete QRT equations studied by Nobe \cite{NobeQRT}. The invariants are
\begin{align*}
I_0 =& \max(Y + \max(A,B,2X), \max(B,X)+ \max(A,B+ X), \\
& \hspace{1cm}2Y + \max(B,X))-B- X -Y,\\
I_1 =& \max(A,B+ X, B+Y, 2X+Y,X +2 Y )- X - Y,\\
I_2 =& \max(A,B+ X, B +Y,2X+ 2Y) - X - Y.
\end{align*}
At this point, we simplify the calculations by setting $A = B = 0$, in which the level sets are depicted in figure \ref{levelsets}.

We now calculate the difference between the left and right derivatives are (left minus the right) where a non-zero discrepancy indicates non-differentiability. These differences for each of the iterates are given by
\[
\begin{array}{c||c|c|c|c|c}
\sigma  & X_3 & X_4 & X_5 & X_6 & X_7 \\ \hline \hline
0 & \{0,1\}_{X_0} & \underline{0} & \underline{0} & 0 & 1 \\
1 & \{-1,0\}_{X_0} & \{-1,0\}_{X_0} & \{1,0\}_{X_0} & \underline{0} & \underline{0} \\
2 & \{-2,-1\}_{X_0} & 1 & \underline{0} & \underline{0} & 0 \\
3 & \{-3,-2\}_{X_0} & 5 & \{-12,-8\}_{X_0} & 19 & \{-45,30\}_{X_0}
\end{array}.
\]
where we have used the notation, $\{d_1,d_2\}_{X_0}$ to mean there is a discrepancy of $d_1$ if $X_0 <0 $ and $d_2$ if $X_0 > 0$ (we omit the case $X_0 = 0$ for concision). There are more complicated formulas for more general $A$ and $B$ that we omit for concision. The point of this exercise is that at some point, we have a pair of $0$ values (underlined above), indicating that the functions are continuous in each of the integrable cases, whereas it is possible to show for $\sigma=3$ that $X_k$ is discontinuous for all $k > 1$.

To understand this pattern, we first consider the system \eqref{QRT1} as being defined over $\mathbb{K}$, with the specialization $a = b = 1$. To consider the tropical singularity confinement of the tropical map, one perturbs the singular initial condition by an additive factor, $\epsilon$. As addition in the max-plus semifield is analogous to multiplication over $\mathbb{K}$, we take a perturbed initial condition of $x_1 = -t^{-\delta}$, which in the case $\sigma =2$ for example, the first few iterates are
\begin{align*}
x_2 =& -\frac{\Delta  t^{\delta }}{x_0},\\
x_3 =& \frac{x_0 \left(\Delta -x_0 t^{-\delta }\right)}{\Delta ^2},\\
x_4 =& \frac{\Delta  \left(x_0^2-\Delta  t^{\delta } \left(\Delta +x_0\right)\right)}{x_0 \left(x_0-\Delta  t^{\delta }\right){}^2},\\
x_5 =& -\frac{t^{\delta
   } \left(\Delta  t^{\delta }-x_0\right) \left(\Delta ^3 t^{\delta }+x_0 \left(\Delta ^2+x_0\right) \left(\Delta  t^{\delta }-x_0\right)\right)}{\left(x_0^2-\Delta  t^{\delta } \left(\Delta +x_0\right)\right){}^2},\\
x_6 =& \frac{x_0 
  \left(-\Delta ^3 x_0 t^{3 \delta }+\Delta ^2 t^{2 \delta } \left(\Delta ^2+3 x_0 \left(x_0+1\right)\right)-\Delta  x_0^2 \left(3 x_0+2\right) t^{\delta }+x_0^4\right)}{t^{2 \delta }\left(\Delta ^3
   t^{\delta }+x_0 \left(\Delta ^2+x_0\right) \left(\Delta  t^{\delta }-x_0\right)\right)^2}\\
   & \times  \left(x_0^2-\Delta  t^{\delta } \left(\Delta +x_0\right)\right)
\end{align*}
where we have introduced notation for the factor
\[
\Delta = 1- t^{\delta},
\]
whose valuation is the discontinuous function
\[
\nu(\Delta) = \left\{ 
\begin{array}{l p{5cm}} 
\max(\delta,0) & for $\delta \neq 0$,\\
-\infty & for $\delta = 0$.
\end{array}\right.
\]
This factor of $\Delta$ encapsulates the correspondence between discontinuities and singularities. If we replace $t$ with a number, the limit as $\delta \to 0$ gives us the usual singularity pattern, and if we leave $t$ as an indeterminate and apply the non-Archimedean valuation, we obtain the pattern of non-differentiability. One of the reasons this works is because $t^{\delta}$ is algebraically a generic arbitrary initial condition, hence, we may assume that lemma \ref{equal} holds. If one expands out the iterates, the factors of $\Delta$ explain precisely the degrees of the non-differentiability. 

If one performs the same test on the case in which $\sigma = 3$, we readily find the iterates over $\mathbb{K}$ are
\begin{align*}
x_2 =& \frac{\Delta  t^{2 \delta }}{x_0},\\
x_3 =& -\frac{x_0^2 t^{-5 \delta } \left(\Delta  t^{2 \delta }+x_0\right)}{\Delta ^3},\\
x_4 =& \frac{\Delta ^5 t^{8 \delta } \left(-\Delta ^3 t^{5 \delta }+\Delta  x_0^2 t^{2 \delta }+x_0^3\right)}{x_0^5 \left(\Delta  t^{2 \delta }+x_0\right){}^3},\\
x_5 =& \frac{x_0^8 \left(\Delta  t^{2 \delta }+x_0\right){}^5 \left(x_0^2 \left(\Delta  t^{2 \delta }+x_0\right) \left(\Delta ^5 t^{8 \delta }+x_0^3 \left(t^{2 \delta }-t^{3 \delta }+x_0\right){}^2\right)-\Delta ^8 t^{13 \delta }\right)}{t^{19 \delta }\Delta
   ^{12} \left(\Delta ^3 t^{5 \delta }-x_0^2 \left(\Delta  t^{2 \delta }+x_0\right)\right){}^3},
\end{align*}
which explains precisely the degree of the non-differentiability observed above. If we allow $t$ to be some constant, the limit as $\delta \to 0$ gives us the usual singularity confinement, whereas, the non-Archimedean valuation gives us the pattern of non-differentiability. Since the singularity is not confined over the field, we will always find a prefactor of $\Delta$ to some power in the iterates, which will give us some non-differentiability of the corresponding ultradiscrete equation. We have simplified the calculations by presenting the case in which there is just one point of-non-differentiability, however, the arguments work equally well for mappings with more points of non-differentiability, such as \eqref{uP3}.

\subsection{Tropical Singularity confinement for P$\Delta$E's}

We first consider singularity confinement, as it is described in \cite{Gramani:DPainleveProperty}. We specialize this to quad-graphs, specified by a multilinear function
\begin{equation}\label{q}
q(w_{l,m},w_{l+1,m},w_{l,m+1},w_{l+1,m+1};\alpha,\beta) = 0,
\end{equation}
meaning that we may solve for any of the lattice variables. For example, in solving for $w_{l+1,m+1}$, we use the expression 
\begin{align*}
q = \dfrac{\partial}{\partial w_{l+1,m+1}} q(w_{l,m},w_{l+1,m}&,w_{l,m+1},w_{l+1,m+1};\alpha,\beta)w_{l+1,m+1}\\
& +  q(w_{l,m},w_{l+1,m},w_{l,m+1},0;\alpha,\beta),
\end{align*}
to give $w_{l+1,m+1}$. If we use this expression to solve in the positive $l$ and $m$ directions, we may take the concept of a singularity to be one in which the initial conditions are such that
\begin{align*}
\dfrac{\partial}{\partial w_{l+1,m+1}} q(w_{l,m},w_{l+1,m},w_{l,m+1},w_{l+1,m+1};\alpha,\beta) &= 0,\\
q(w_{l,m},w_{l+1,m},w_{l,m+1},0;\alpha,\beta) &= 0.
\end{align*}
The key property is having the first equation hold, which admits two interpretations in the literature:
\begin{itemize}
\item{The second equation does not necessarily hold and $w_{l+1,m+1}$ involves a denominator that becomes $0$, hence, the value of $w_{l+1,m+1}$ tends to $\infty$.}
\item{The second equation necessarily holds, hence, $w_{l+1,m+1}$ may be chosen arbitrarily.} 
\end{itemize}
In both cases, note that there is some loss of information, as $w_{l+1,m+1}$ does not contain information regarding $w_{l,m}$. We acknownledge the second interpretation, as considered by Atkinson \cite{Atkinson:Singularities}, however, for this work we adopt the first interpretation, which is consistent with the original implementation of singularity confinement \cite{Gramani:DPainleveProperty}.

\begin{defn}
A singularity is confined if one may iterate beyond singular boundaries to reclaim initial conditions in some limit.
\end{defn}

\begin{figure}[!ht]
\begin{tikzpicture}[scale=.8]
\draw[step=2.0] (-2.5,-2.5) grid (2.5,2.5);
\begin{scope}[yshift=.3cm,xshift=.4cm]
	\node at (0,0) {$1$};
	\node at (-2,0) {$x_1$};
	\node at (-3.4,2) {$-\frac{\beta}{\alpha}$};
	\node at (0,-2) {$x_2$};
	\node at (2.3,-2) {$-\frac{\alpha}{\beta}$};
	\node at (2,0) {$\infty$};
	\node at (0,2) {$\infty$};
	\node at (2.7,2) {$\frac{\beta  x_1-\alpha  x_2}{\alpha  x_1-\beta  x_2}$};
\end{scope}
\foreach \x in {-2,0,...,2}
{
	\foreach \y in {-2,0,...,2}
	{
		\filldraw[blue] (\x,\y) circle (.1);
	}
}
\end{tikzpicture}
\caption{The singularity pattern of the lattice Korteweg-de Vries equation. \label{singmKdV}}
\end{figure}
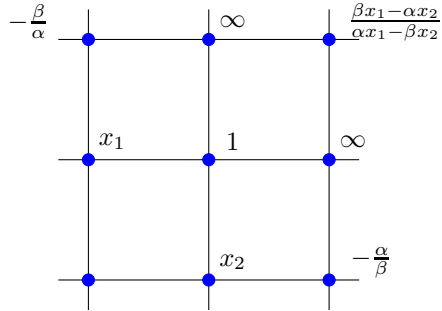

Let us consider the example of the modified Korteweg-de Vries equation \cite{Hirota:dSG, Nijhoff:lkdvreview}, or $H3(\delta= 0)$ as it appears in \cite{ABS:ListI, ABS:ListII} (with the parameter $\beta \to -\beta$),  
\begin{align}\label{H3}
\alpha(w_{l,m}w_{l+1,m} - w_{l,m+1}w_{l+1,m+1}) + \beta(w_{l,m}w_{l,m+1} - w_{l+1,m}w_{l+1,m+1}) =0,
\end{align}
which we solve in $w_{l+1,m+1}$ to obtain 
\begin{equation}\label{mKdV}
w_{l+1,m+1} = w_{l,m} \dfrac{\alpha w_{l+1,m}  + \beta w_{l,m+1}}{\beta w_{l+1,m} + \alpha w_{l,m+1}}.
\end{equation}
Differentiating with respect to $w_{l+1,m+1}$ shows us that the equation is clearly singular when $w_{l+1,m} = -\alpha w_{l,m+1}/\beta$, hence, let us define initial conditions 
\begin{align*}
w_{0,0}&= 1+\epsilon, \hspace{2cm} w_{-1,1} = -\dfrac{\beta}{\alpha}, \\ 
w_{-1,0} &= x_1,			\hspace{2cm} w_{-1,1} = -\dfrac{\alpha}{\beta}, \\
w_{0,-1} &= x_2.
\end{align*}
From these initial conditions, it is clear that $w_{1,0}$ and $w_{0,1}$ tend to $\infty$ as $\epsilon \to 0$. This configuration is depicted in figure \ref{singmKdV}.

The key observations is that $w_{1,0} \to \infty$ and $w_{0,1} \to \infty$ as $\epsilon \to 0$ the limit of $w_{1,1}$ as $\epsilon \to 0$ involves both $x_1$ and $x_2$. As in the ordinary differential setting, this singularity pattern may also be observed by considering a multiplicative deviation from the singular boundary. That is, if we replace $w_{0,0} = e^{\delta}$, the limit as $\delta \to 0$ gives us the same pattern. 

\begin{rem}
Another way in which we may approach the singular boundary is if we were to let $w_{-1,1} = -\dfrac{\alpha}{\beta} + \epsilon_1$ and $w_{1,-1} = -\dfrac{\beta}{\alpha} + \epsilon_2$. If $\epsilon_1 = a\epsilon$ and $\epsilon_2 = b \epsilon$, the limit of $w_{1,1}$ as $\epsilon \to 0$ depends on $a/b$. In other words, $w_{1,1}$ depends on the blow-up of $(w_{-1,1},w_{1,-1})$ at $\left(-\dfrac{\alpha}{\beta},-\dfrac{\beta}{\alpha}\right)$.
\end{rem}

If we now consider ultradiscrete lattice equations, there are many that have appeared in the literature, such as the ultradiscrete modified Korteweg-de Vries equation \cite{udmKdV:BBS, Quispel:UDLax}, ultradiscrete Lotka Volterra equation \cite{ultradiscretization} and the ultradiscrete Kadomtsev-Petviashvili equation \cite{udKP:BBS}. Our examples will be from a particular class of equations, which we would consider the ultradiscrete analogue of those in the ABS-list \cite{ABS:ListI}. We may now interpret \eqref{q} as a function over $\mathbb{K}$, then the image, under the valuation, is some tropically multi-affine linear functions
\[
Q(W_{l,m},W_{l+1,m}, W_{l,m+1}, W_{l+1,m+1}; A, B).
\]
Given the correspondence between $0$'s and points of non-differentiability, we may interpret equality as being when the left hand side being non-differentiable. We claim that this is a valid tropical analogue of interpreting a singularity.

If we attempt to find where this function is not differentiable for one of its' arguments, say $W_{l+1,m+1}$, it is abundantly clear that $Q$ admits a representation of the form
\begin{align*}
Q &= \max( W_{l+1,m+1} + F(W_{l,m}, W_{l+1,m}, W_{l,m+1}), G(W_{l,m}, W_{l+1,m}, W_{l,m+1})).
\end{align*}
It is clear that $Q$ is not differentiable in $W_{l+1,m+1}$ when 
\[
W_{l+1,m+1} = G(W_{l,m}, W_{l+1,m}, W_{l,m+1}) - F(W_{l,m}, W_{l+1,m}, W_{l,m+1}).
\]
If $W_{l,m}$, $W_{l+1,m}$ and $W_{l,m+1}$ are chosen generically, this evolution makes sense. If $W_{l,m}$, $W_{l+1,m}$ and $W_{l,m+1}$ is on the set (of measure $0$) in which $F$ or $G$ are not differentiable, $W_{l+1,m+1}$, the evolution in terms of $G$ and $F$ makes sense if we consider differentiability in the variable $W_{l+1,m+1}$, however, strictly speaking, if we just require non-differentiability of $Q$ in any direction, $W_{l+1,m+1}$ may be chosen to be any value such that either $F + W_{l+1,m+1}$ or $G$ dominate the max-expression, making it a true analogue of a singularity.

\begin{defn}
A tropical singularity is confined if, by iterating past non-differentiable boundaries, one reclaims differentiability. 
\end{defn}

Let us take the image of the \eqref{H3}, which, if the parameters are taken to be generic, results in the equation
\begin{align*}
Q =& \max(A + \max(W_{l,m}+ W_{l+1,m}, W_{l,m+1}+W_{l+1,m+1}) ,\\
&\hspace{1.5cm} B+ \max(W_{l,m}+W_{l,m+1}, W_{l+1,m}+W_{l+1,m+1})).
\end{align*}
If we are to now write this in a manner in which the point of non-differentiability in the co-ordinate $W_{l+1,m+1}$ is made clear, we write $Q$ as
\begin{align*}
Q =& \max(W_{l+1,m+1} + \max( A + W_{l,m+1}, B + W_{l+1,m}), \\
&\hspace{1.5cm} W_{l,m} + \max(A + W_{l+1,m}, B+W_{l,m+1})),
\end{align*}
hence, the ultradiscrete evolution is given by
\begin{align}
W_{l+1,m+1} = W_{l,m} &+ \max(A + W_{l+1,m}, B+W_{l,m+1}) \label{udmKdV}\\
&- \max( A + W_{l,m+1}, B + W_{l+1,m})\nonumber,
\end{align}
which has appeared in the literature \cite{udmKdV:BBS, Quispel:UDLax, Ormerod:qP6}. It is easy to show, and interesting to note, that this equation is consistent around a cube in the usual sense. Let us now consider the analogous initial conditions, 
\begin{align*}
W_{0,0} &= \epsilon, \hspace{2cm} w_{-1,1} = B-A,\\ 
W_{-1,0} &= X_1,\hspace{2cm} w_{-1,1} = A-B, \\
W_{0,-1} &= X_2,
\end{align*}
which are designed to be singular in the sense that the functions are non-differentiable in the co-ordinates comprised of the staircase of initial conditions. We readily find, using \eqref{udmKdV}, the values
\begin{align*}
W_{0,1} =& X_1 + B+A+\max(0,\epsilon) -\max(2B,2A+\epsilon),\\ 
W_{1,0} =& X_2 + B+A+\max(0,\epsilon) -\max(2A,2B+\epsilon),\\
W_{1,1} =& -\epsilon + \max( \epsilon+\max(3A+X_2,3B+X_1), A+B+\max(A + X_1+B+X_2))\\
&- \max(  \max( 3A+X_1,3B+X_2), A+B + \epsilon +\max(A+X_2,B+X_1)).
\end{align*}
The first thing to notice is that $W_{0,1}$ and $W_{1,0}$ clearly have discontinuities in the derivative with respect to $\epsilon$ as $\epsilon \to 0$. While the last expression is somewhat complicated, so long as $A+X_1 \neq B+X_2$ or $A+X_2 \neq B+X_1$, this expression does not depend on the direction of the limit as $\epsilon \to 0$, hence, the derivative, under these conditions, is continuous. 

Now let us return to \eqref{H3}, but define the variables to be elements of $\mathbb{K}$, such that $\nu(\alpha) = A$ and $\nu(\beta) = B$. We now let $x_{0,0}$ be $t^{-\delta}$, where the iterates are
\begin{align*}
w_{0,1} =& \frac{x_1 \left(\beta ^2 t^{\delta }-\alpha ^2\right)}{\alpha  \beta  \Delta },\\
w_{1,0} =& \frac{x_2 \left(\alpha ^2 t^{\delta }-\beta ^2\right)}{\alpha  \beta  \Delta },\\
w_{1,1} =& \frac{t^{-\delta } \left(t^{\delta } \left(\alpha ^3 x_2+\beta ^3 x_1\right)-\alpha  \beta  \left(\alpha  x_1+\beta  x_2\right)\right)}{\alpha  \beta  t^{\delta }
   \left(\alpha  x_2+\beta  x_1\right)-\alpha ^3 x_1-\beta ^3 x_2}.
\end{align*}
Once again the factor, $\Delta= 1-t^{\delta}$, is the term that encapsulates both the forms of singularity confinement. If we let $\delta \to 0$, where $t$ is chosen to be some number, then we reclaim singularity confinement as it appeared originally \cite{Gramani:DPainleveProperty}. On the other hand, if we apply our valuation, $\nu(\Delta)$ is the factor that appears and disappears in accordance with the observed ultradiscrete singularity confinement. 

\section{Conclusion}

There have been a number of recent results that have reiterated the importance of the integrability of ultradiscrete equations. The Lax integrability of an ultradiscrete analogue of the sixth Painlev\'e equation \cite{Ormerod:qP6}, the existence of ultradiscrete hypergeometric solutions \cite{Ormerod:uhypergeometric}, the interpretation of the ultradiscrete Quispel-Roberts-Thompson system \cite{NobeQRT} and importantly, the connection between ultradiscrete equations and entropy, as observed by Fordy and Hone \cite{fordy2012discrete}. 

Another very interesting link, although not immediately obvious, is the recent arithmetic version of singularity confinement over the rational numbers, using $p$-adic valuations \cite{DPIIFiniteField}. The $p$-adic's are also non-Archimedean valuation fields, and much of the dynamics of these integrable systems under the image of the valuation of the $p$-adics are similar. 

In short, through a proper understanding of what the ultradiscrete analogue of singularities are, through the interpretation of an ultradiscretization of arbitrary rational functions, we have at our disposal the powerful tool of tropical algebraic geometry. With the ripening of tropical geometry, I believe there will be a renewed interest in geometry of ultradiscrete equations on the horizon, and this result presents a way of understanding ultradiscrete equations in this context.

\section*{Acknowledgments}

We would like to thank Professor Nalini Joshi and Professor Kenji Kajiwara for their encouragement in this work. This research is supported by Australian Research Council Discovery Grant \#DP110100077.

\bibliography{C:/Mathematics/TeX/refs}{}

\def\cprime{$'$}
\begin{thebibliography}{10}

\bibitem{ABS:ListI}
V.~E. Adler, A.~I. Bobenko, and Yu.~B. Suris.
\newblock Classification of integrable equations on quad-graphs. {T}he
  consistency approach.
\newblock {\em Comm. Math. Phys.}, 233(3):513--543, 2003.

\bibitem{ABS:ListII}
V.~E. Adler, A.~I. Bobenko, and Yu.~B. Suris.
\newblock Discrete nonlinear hyperbolic equations: classification of integrable
  cases.
\newblock {\em Funct. Anal. Appl.}, 43(1):3--17, 2009.

\bibitem{Atkinson:Singularities}
James Atkinson.
\newblock Singularities of type-{Q} {ABS} equations.
\newblock {\em SIGMA Symmetry Integrability Geom. Methods Appl.}, 7:Paper 073,
  14, 2011.

\bibitem{Viallet}
M.~P. Bellon and C.-M. Viallet.
\newblock Algebraic entropy.
\newblock {\em Comm. Math. Phys.}, 204(2):425--437, 1999.

\bibitem{BieriGroves}
Robert Bieri and J.~R.~J. Groves.
\newblock The geometry of the set of characters induced by valuations.
\newblock {\em J. Reine Angew. Math.}, 347:168--195, 1984.

\bibitem{Duistermaat}
Johannes~J. Duistermaat.
\newblock {\em Discrete integrable systems}.
\newblock Springer Monographs in Mathematics. Springer, New York, 2010.

\bibitem{Amoebas}
Manfred Einsiedler, Mikhail Kapranov, and Douglas Lind.
\newblock Non-{A}rchimedean amoebas and tropical varieties.
\newblock {\em J. Reine Angew. Math.}, 601:139--157, 2006.

\bibitem{fordy2012discrete}
A.~Fordy and A.~Hone.
\newblock Discrete integrable systems and poisson algebras from cluster maps.
\newblock {\em arXiv preprint arXiv:1207.6072}, 2012.

\bibitem{Gramani:DPainleveProperty}
B.~Grammaticos, A.~Ramani, and V.~Papageorgiou.
\newblock Do integrable mappings have the {P}ainlev\'e property?
\newblock {\em Phys. Rev. Lett.}, 67(14):1825--1828, 1991.

\bibitem{Gramani:ABSSC}
Basil Grammaticos and Alfred Ramani.
\newblock Singularity confinement property for the (non-autonomous)
  {A}dler-{B}obenko-{S}uris integrable lattice equations.
\newblock {\em Lett. Math. Phys.}, 92(1):33--45, 2010.

\bibitem{Hirota:dSG}
Ryogo Hirota.
\newblock Nonlinear partial difference equations. {III}. {D}iscrete
  sine-{G}ordon equation.
\newblock {\em J. Phys. Soc. Japan}, 43(6):2079--2086, 1977.

\bibitem{s-ultradiscretization}
S.~Isojima, B.~Grammaticos, A.~Ramani, and J.~Satsuma.
\newblock Ultradiscretization without positivity.
\newblock {\em J. Phys. A}, 39(14):3663--3672, 2006.

\bibitem{Joshi:TropSC}
N.~Joshi and S.~Lafortune.
\newblock Integrable ultra-discrete equations and singularity analysis.
\newblock {\em Nonlinearity}, 19(6):1295--1312, 2006.

\bibitem{Ormerod:uP3}
N.~Joshi, F.~W. Nijhoff, and C.~Ormerod.
\newblock Lax pairs for ultra-discrete {P}ainlev\'e cellular automata.
\newblock {\em J. Phys. A}, 37(44):L559--L565, 2004.

\bibitem{Ormerod:theory}
Nalini Joshi and Chris Ormerod.
\newblock The general theory of linear difference equations over the max-plus
  semi-ring.
\newblock {\em Stud. Appl. Math.}, 118(1):85--97, 2007.

\bibitem{DPIIFiniteField}
M~Kanki, J~Mada, K~M Tamizhmani, and T~Tokihiro.
\newblock Discrete {P}ainlev\'e {II} equation over finite fields.
\newblock {\em Journal of Physics A: Mathematical and Theoretical},
  45(34):342001, 2012.

\bibitem{UDconserved}
Masataka Kanki, Jun Mada, and Tetsuji Tokihiro.
\newblock Conserved quantities and generalized solutions of the ultradiscrete
  {K}d{V} equation.
\newblock {\em J. Phys. A}, 44(14):145202, 13, 2011.

\bibitem{Lafortune:negativity}
Alex Kasman and St{\'e}phane Lafortune.
\newblock When is negativity not a problem for the ultradiscrete limit?
\newblock {\em J. Math. Phys.}, 47(10):103510, 16, 2006.

\bibitem{FieldforTG}
T.~Markwig.
\newblock A field of generalised {P}uiseux series for tropical geometry.
\newblock {\em Rend. Semin. Mat. Univ. Politec. Torino}, 68(1):79--92, 2010.

\bibitem{Nijhoff:lkdvreview}
Frank~W. Nijhoff and Hans Capel.
\newblock The discrete {K}orteweg-de {V}ries equation.
\newblock {\em Acta Appl. Math.}, 39(1-3):133--158, 1995.
\newblock KdV '95 (Amsterdam, 1995).

\bibitem{NobeQRT}
Atsushi Nobe.
\newblock Ultradiscrete {QRT} maps and tropical elliptic curves.
\newblock {\em J. Phys. A}, 41(12):125205, 12, 2008.

\bibitem{Ochiai}
Tomoshiro Ochiai and Jose~C. Nacher.
\newblock Inversible max-plus algebras and integrable systems.
\newblock {\em J. Math. Phys.}, 46(6):063507, 17, 2005.

\bibitem{Ormerod:uhypergeometric}
Christopher~M. Ormerod.
\newblock Hypergeometric solutions to an ultradiscrete {P}ainlev\'e equation.
\newblock {\em J. Nonlinear Math. Phys.}, 17(1):87--102, 2010.

\bibitem{Ormerod:qP6}
Christopher~M. Ormerod.
\newblock Reductions of lattice m{K}d{V} to $q$-$\mathrm{P}_{VI}$.
\newblock {\em Phys. Lett. A}, 2012.
\newblock In Press.

\bibitem{pin1998tropical}
J.E. Pin.
\newblock Tropical semirings.
\newblock {\em Idempotency (Bristol, 1994)}, pages 50--69, 1998.

\bibitem{Quispel:UDLax}
G.~R.~W. Quispel, H.~W. Capel, and J.~Scully.
\newblock Piecewise-linear soliton equations and piecewise-linear integrable
  maps.
\newblock {\em J. Phys. A}, 34(11):2491--2503, 2001.
\newblock Kowalevski Workshop on Mathematical Methods of Regular Dynamics
  (Leeds, 2000).

\bibitem{Gramani:DiscretePs}
A.~Ramani, B.~Grammaticos, and J.~Hietarinta.
\newblock Discrete versions of the {P}ainlev\'e equations.
\newblock {\em Phys. Rev. Lett.}, 67(14):1829--1832, 1991.

\bibitem{ultimate}
A.~Ramani, D.~Takahashi, B.~Grammaticos, and Y.~Ohta.
\newblock The ultimate discretisation of the {P}ainlev\'e equations.
\newblock {\em Phys. D}, 114(3-4):185--196, 1998.

\bibitem{TropicalGeometry}
J{\"u}rgen Richter-Gebert, Bernd Sturmfels, and Thorsten Theobald.
\newblock First steps in tropical geometry.
\newblock In {\em Idempotent mathematics and mathematical physics}, volume 377
  of {\em Contemp. Math.}, pages 289--317. Amer. Math. Soc., Providence, RI,
  2005.

\bibitem{Sakai:Rational}
Hidetaka Sakai.
\newblock Rational surfaces associated with affine root systems and geometry of
  the {P}ainlev\'e equations.
\newblock {\em Comm. Math. Phys.}, 220(1):165--229, 2001.

\bibitem{udmKdV:BBS}
Daisuke Takahashi and Junta Matsukidaira.
\newblock Box and ball system with a carrier and ultradiscrete modified {K}d{V}
  equation.
\newblock {\em J. Phys. A}, 30(21):L733--L739, 1997.

\bibitem{udKP:BBS}
T.~Tokihiro, D.~Takahashi, and J.~Matsukidaira.
\newblock Box and ball system as a realization of ultradiscrete nonautonomous
  {KP} equation.
\newblock {\em J. Phys. A}, 33(3):607--619, 2000.

\bibitem{ultradiscretization}
T.~Tokihiro, D.~Takahashi, J.~Matsukidaira, and J.~Satsuma.
\newblock From soliton equations to integrable cellular automata through a
  limiting procedure.
\newblock {\em Physical Review Letters}, 76(18):3247--3250, 1996.

\bibitem{YamadaEreps}
Yasuhiko Yamada.
\newblock Tropical affine {W}eyl group representation of type ${E}^{(1)}_n$.
\newblock 2004.

\end{thebibliography}
\bibliographystyle{plain}

\end{document}